# Absence of the non-uniqueness problem of the Dirac theory in a curved spacetime
# Spin-rotation coupling is not physically relevant

M.V. Gorbatenko, V.P. Neznamov[1]

RFNC-VNIIEF, 37 Mira Ave., Sarov, 607188 Russia

Abstract

As opposed to Arminjon's statements, in this work we again assert the absence of the non-uniqueness problem of the Dirac theory in a curved and flat spacetime and illustrate this with a number of examples. Dirac Hamiltonians in arbitrary, including time-dependent, gravitational fields uniquely determine physical characteristics of quantum-mechanical systems irrespective of the choice of the tetrad fields.

Direct spin-rotation coupling that occurs with a certain choice of tetrads does not manifest itself in final physical characteristics of the systems and therefore does not represent a physically relevant effect.

---

[1] E-mail: neznamov@vniief.ru

In the recent time, publications have emerged again [1] - [3], which declare and provide grounds for the assertion that the Dirac theory is non-unique in a curved and even flat spacetime. The proof is based on the demonstration that the form of Dirac Hamiltonians depends on the choice of tetrads. In our opinion, this is absolutely insufficient. To demonstrate the non-equivalence of Dirac Hamiltonians, one should find the difference in physical characteristics of a system under consideration with different choices of tetrads. Such characteristics may include energy spectra of Hamiltonians, mean values of physical quantities, various transition amplitudes and so on.

We share the conclusions of previous studies [4], [5] on the independence of physical characteristics of the Dirac theory on the choice of tetrads.

In [6] - [8], using the methods of pseudo-Hermitian quantum mechanics [9] - [11] for arbitrary, including time-dependent, gravitational fields, we developed an algorithm to transform any Dirac Hamiltonian in a curved spacetime with an arbitrary choice of tetrads into the $\eta$ - representation, in which the Hamiltonian becomes self-conjugate, and the scalar product of wave functions becomes flat. The choice of different tetrads for the same physical system can lead in the $\eta$ - representation to different forms of self-conjugate Hamiltonians. However, they will always be related by unitary transformations associated with spacetime rotations of Dirac matrices. It is evident that such Hamiltonians are physically equivalent. The choice of tetrads by a researcher is governed by convenience considerations.

One can handle Dirac Hamiltonians in a curved spacetime using Parker's weight operator in the scalar product of wave functions [12], or treat them in the $\eta$ - representation with a flat scalar product, using the common apparatus of quantum mechanics. In both cases physical characteristics of the systems remain identical.

Let us give some examples to illustrate this.

Below we use the system of units $\hbar = c = G = 1$, where $\hbar$ is the Planck constant, $c$ is the speed of light, and $G$ is the gravitational constant.

For the first three examples we use the signature

$$\eta_{\underline{\alpha\beta}} = diag[-1,1,1,1]. \tag{1}$$

Local indices are underlined, and global indices are not underlined. Hence, for Dirac $\gamma$ - matrices,

$$\gamma^{\underline{\alpha}}\gamma^{\underline{\beta}} + \gamma^{\underline{\beta}}\gamma^{\underline{\alpha}} = 2\eta^{\underline{\alpha\beta}}E \tag{2}$$

$$\gamma^{\alpha}\gamma^{\beta} + \gamma^{\beta}\gamma^{\alpha} = 2g^{\alpha\beta}E. \tag{3}$$

In (2), (3), $E$ is a 4 x 4 unity matrix.

Tetrads are defined by the relationships

$$H_{\underline{\alpha}}^{\mu} H_{\underline{\beta}}^{\nu} g_{\mu\nu} = \eta_{\underline{\alpha}\underline{\beta}}. \tag{4}$$

The relationship between $\gamma^\alpha$ and $\gamma^{\underline{\alpha}}$ is given by the equalities

$$\gamma^\alpha = H_{\underline{\beta}}^{\alpha} \gamma^{\underline{\beta}}. \tag{5}$$

Parker's weight operator equals

$$\rho = \sqrt{-g}\,\gamma_{\underline{0}}\gamma^0. \tag{6}$$

Example 1.

In [6], three Hamiltonians, corresponding to three tetrad fields, and a self-conjugate Hamiltonian in the $\eta$- representation are derived for a weak Kerr field.

a) Killing tetrad field

$$\begin{aligned}
H_k = &\, im\gamma_{\underline{0}} - im\frac{M}{R}\gamma_{\underline{0}} - i\gamma_{\underline{0}}\gamma^{\underline{k}}\frac{\partial}{\partial x^k} + 2i\frac{M}{R}\gamma_{\underline{0}}\gamma^{\underline{k}}\frac{\partial}{\partial x^k} + \\
&+ \frac{i}{2}\frac{MR_k}{R^3}\gamma_{\underline{0}}\gamma^{\underline{k}} + 2i\frac{M(J_{kl}R_l)}{R^3}\frac{\partial}{\partial x^k} - 2im\frac{M(J_{kl}R_l)}{R^3}\gamma^{\underline{k}} + \\
&+ 2i\frac{M(J_{ml}R_l)}{R^3}S_{\underline{mk}}\frac{\partial}{\partial x^k} - \frac{i}{2}\left\{\frac{M}{R^3}J_k - 3\frac{M(J_l R_l)R_k}{R^5}\right\}\gamma_{\underline{5}}\gamma_{\underline{0}}\gamma_{\underline{k}}.
\end{aligned} \tag{7}$$

$$\rho = 1 + \frac{3M}{R} + 2\frac{M(J_{km}R_m)}{R^3}\gamma_{\underline{0}}\gamma_{\underline{k}}; \tag{8}$$

b) tetrad field in symmetric gauge

$$\begin{aligned}
H_s = &\, im\gamma_{\underline{0}} - i\gamma_{\underline{0}}\gamma_{\underline{k}}\frac{\partial}{\partial x^k} - im\frac{M}{R}\gamma_{\underline{0}} + 2i\frac{M}{R}\gamma_{\underline{0}}\gamma^{\underline{k}}\frac{\partial}{\partial x^k} + \frac{i}{2}\frac{MR_k}{R^3}\gamma_{\underline{0}}\gamma^{\underline{k}} + \\
&+ 2i\frac{M(J_{kl}R_l)}{R^3}\frac{\partial}{\partial x^k} - im\frac{M(J_{kl}R_l)}{R^3}\cdot\gamma^{\underline{k}} + i\frac{M(J_{ml}R_l)}{R^3}S_{\underline{mk}}\frac{\partial}{\partial x^k}.
\end{aligned} \tag{9}$$

$$\rho = 1 + \frac{3M}{R} + \frac{MJ_{km}R_m}{R^3}\gamma_{\underline{0}}\gamma_{\underline{k}}; \tag{10}$$

c) tetrad field of Hehl and Ni [13]

$$\begin{aligned}
H_{H-N} = &\, im\gamma_{\underline{0}} - im\frac{M}{R}\gamma_{\underline{0}} - i\gamma_{\underline{0}}\gamma_{\underline{k}}\frac{\partial}{\partial x^k} + 2i\frac{M}{R}\gamma_{\underline{0}}\gamma^{\underline{k}}\frac{\partial}{\partial x^k} + \frac{i}{2}\frac{MR_k}{R^3}\gamma_{\underline{0}}\gamma^{\underline{k}} + \\
&+ 2i\frac{M(J_{kl}R_l)}{R^3}\frac{\partial}{\partial x^k} + \frac{i}{2}\left\{\frac{M}{R^3}J_k - 3\frac{M(J_l R_l)R_k}{R^5}\right\}\gamma_{\underline{5}}\gamma_{\underline{0}}\gamma_{\underline{k}}.
\end{aligned} \tag{11}$$

$$\rho = 1 + \frac{3M}{R}; \tag{12}$$

d) self-conjugate Hamiltonian in the $\eta$- representation

$$H_\eta = im\gamma_0 - im\frac{M}{R}\gamma_0 - i\gamma_0\gamma_k\frac{\partial}{\partial x^k} + 2i\frac{M}{R}\gamma_0\gamma^k\frac{\partial}{\partial x^k} - i\frac{MR_k}{R^3}\gamma_0\gamma^k +$$
$$+ 2i\frac{M(J_{kl}R_l)}{R^3}\frac{\partial}{\partial x^k} + \frac{i}{2}\left\{\frac{M}{R^3}J_k - 3\frac{M(J_l R_l)R_k}{R^5}\right\}\gamma_5\gamma_0\gamma_k. \tag{13}$$

$$\rho = 1. \tag{14}$$

In (7) – (14), $M$ is the mass of a source of the Kerr gravitational field, $J_{km}$ is the angular momentum tensor of the Kerr field, $S_{mk} = \frac{1}{2}(\gamma_m\gamma_k - \gamma_k\gamma_m)$.

Each of the Hamiltonians (7), (9), (11), (13) differ from each other in their form. However, with the transition to the $\eta$- representation, all the Hamiltonians become the same, which proves their physical equivalence.

Example 2.

We know that a free Dirac Hamiltonian in spherical coordinates of Minkowski space can be written in two ways resulting in substantially different expressions (see, e.g., [14])

$$H_1 = im\gamma_0 - i\gamma_0\left\{\gamma_1\left(\frac{\partial}{\partial r} + \frac{1}{r}\right) + \frac{1}{r}\gamma_2\left(\frac{\partial}{\partial\theta} + \frac{1}{2}\text{ctg}\theta\right) + \frac{1}{r\sin\theta}\gamma_3\frac{\partial}{\partial\varphi}\right\}, \tag{15}$$

$$H_2 = im\gamma_0 - i\gamma_0\left\{\gamma_r\frac{\partial}{\partial r} + \gamma_\theta\frac{1}{r}\frac{\partial}{\partial\theta} + \gamma_\varphi\frac{1}{r\sin\theta}\frac{\partial}{\partial\varphi}\right\}. \tag{16}$$

In (16),

$$\gamma_r = \sin\theta[\gamma_1\cos\varphi + \gamma_2\sin\varphi] + \gamma_3\cos\theta = R\gamma_1 R^{-1}$$
$$\gamma_\theta = \cos\theta[\gamma_1\cos\varphi + \gamma_2\sin\varphi] - \gamma_3\sin\theta = R\gamma_2 R^{-1} \tag{17}$$
$$\gamma_\varphi = -\gamma_1\sin\varphi + \gamma_2\cos\varphi = R\gamma_3 R^{-1}.$$

The set $\{\gamma_r, \gamma_\theta, \gamma_\varphi\}$ is related to the set $\{\gamma_1, \gamma_2, \gamma_3\}$ by a unitary matrix $R$,

$$R = R_1 T_1 R_2 T_2$$
$$R_1 = \exp\left(-\frac{\varphi}{2}\gamma_1\gamma_2\right); \quad T_1 = \frac{1}{\sqrt{2}}\gamma_5\gamma_1(E + \gamma_1\gamma_2) \tag{18}$$
$$R_2 = \exp\left(-\frac{\theta}{2}\gamma_2\gamma_3\right); \quad T_2 = \frac{1}{\sqrt{2}}\gamma_5\gamma_2(E + \gamma_3\gamma_1).$$

This shows that the Hamiltonians (15), (16) are physically equivalent, because they are related by the unitary transformation (18)

$$H_2 = RH_1 R^{-1}, \quad R^{-1} = R^+. \tag{19}$$

Example 3.

In [8], the following form of a Dirac Hamiltonian in Boyer-Lindquist coordinates is derived for a weak Kerr field:

$$H_{B-L} = im\left(1 - \frac{r_0}{2r}\right)\gamma_{\underline{0}} - i\left(1 - \frac{r_0}{r}\right)\gamma_{\underline{0}}\gamma_{\underline{1}}\left(\frac{\partial}{\partial r} + \frac{1}{r}\right) -$$
$$- i\left(1 - \frac{r_0}{2r}\right)\frac{1}{r}\left[\gamma_{\underline{0}}\gamma_{\underline{2}}\left(\frac{\partial}{\partial \theta} + \frac{1}{2}\operatorname{ctg}\theta\right) + \gamma_{\underline{0}}\gamma_{\underline{3}}\frac{1}{\sin\theta}\frac{\partial}{\partial \varphi}\right] - \quad (20)$$
$$- \gamma_{\underline{0}}\gamma_{\underline{1}}\frac{r_0}{2r^2} - i\frac{r_0 a}{r^3}\frac{\partial}{\partial \varphi} - i\frac{3}{4}\frac{r_0 a}{r^3}\gamma_{\underline{3}}\gamma_{\underline{1}}\sin\theta.$$

Let us compare this Hamiltonian with Hamiltonian (13). We rewrite (13) using somewhat different notation

$$H_\eta = im\left(1 - \frac{r_0}{2r}\right)\gamma_{\underline{0}} - i\left(1 - \frac{r_0}{r}\right)\gamma_{\underline{0}}\gamma_{\underline{k}}\frac{\partial}{\partial x^k} -$$
$$- i\frac{r_0}{2r^3}\gamma_{\underline{0}}\gamma_{\underline{k}}x_k - i\frac{r_0 a}{r^3}\left(x_1\frac{\partial}{\partial x_2} - x_2\frac{\partial}{\partial x_1}\right) + \quad (21)$$
$$+ i\frac{r_0 a}{4r^3}\left[\gamma_{\underline{1}}\gamma_{\underline{2}}\left(1 - 3\frac{x_3^2}{r^2}\right) - \gamma_{\underline{2}}\gamma_{\underline{3}}\frac{3x_3 x_1}{r^2} - \gamma_{\underline{3}}\gamma_{\underline{1}}\frac{3x_3 x_2}{r^2}\right].$$

In (21), $r_0 = 2M$, $\mathbf{J} = M\mathbf{a}$, $\mathbf{a} = (0, 0, a)$.

In (20), (21), the summands without momentum $a$ correspond to the Schwarzschild solution. In (21), these summands are written in Cartesian coordinates, and in (20), in spherical coordinates, to which Boyer-Lindquist coordinates reduce in the weak-field approximation. These terms in (20), (21) are physically equivalent to each other.

The summands with the momentum of rotation $a$ in (20), (21) differ substantially from each other. However, in [8], these terms in (20), (21) are shown to be also physically equivalent using matrices (17), (18).

In the examples below we use a modification of signature (1)

$$\eta_{\alpha\beta} = diag[1, -1, -1, -1]. \quad (22)$$

Example 4.

For the solution

$$ds^2 = V^2(\mathbf{x})dt^2 - W^2(\mathbf{x})d\mathbf{x}^2 \quad (23)$$

Obukhov [15] obtained a self-conjugate Hamiltonian with a flat scalar product of wave functions

$$H_{ob} = \beta mV + \frac{1}{2}\left[\boldsymbol{\alpha}\mathbf{p}\frac{V}{W} + \frac{V}{W}\boldsymbol{\alpha}\mathbf{p}\right] \qquad (24)$$

In (24), $\beta = \gamma^{\underline{0}}$, $\alpha^k = \gamma^{\underline{0}}\gamma^{\underline{k}}$.

Then, after the unitary Eriksen-Kolsrud transformation [16], in the approximation of a weak gravitational field, Hamiltonian (24) becomes equal to

$$H_{E-K} = \beta\left(mV + \frac{\mathbf{p}^2}{2m}\right) - \frac{\beta}{4m}\{p^2, V-1\} + \\ + \frac{\beta}{2m}\left\{\mathbf{p}^2, \frac{V}{W}-1\right\} + \frac{\beta}{4m}\left[2\boldsymbol{\Sigma}(\mathbf{f}\times\mathbf{p}) + \nabla\mathbf{f}\right] + \frac{1}{2}(\boldsymbol{\Sigma}\boldsymbol{\Phi}). \qquad (25)$$

In (25), $\boldsymbol{\Phi} = \nabla V$; $f = \nabla\left(\frac{V}{W}\right)$; $\boldsymbol{\Sigma} = \begin{pmatrix}\boldsymbol{\sigma} & 0 \\ 0 & \boldsymbol{\sigma}\end{pmatrix}$.

The last summand in (25) can be interpreted as direct interaction between the spin of a Dirac particle and gravitation.

However, for correct classical interpretation of individual summands in the Hamiltonian, initial expression (24) should be subjected to a unitary Foldy-Wouthuysen transformation [17], [18], [19].

As a result, A.Silenko and O. Teryaev [18] obtained the following expression for the transformed Hamiltonian:

$$H_{FW} = \beta\left(mV + \frac{\mathbf{p}^2}{2m}\right) - \frac{\beta}{4m}\{p^2, V-1\} + \frac{\beta}{2m}\left\{\mathbf{p}^2, \frac{V}{W}-1\right\} + \\ + \frac{\beta}{4m}\left[2\boldsymbol{\Sigma}(\mathbf{f}\times\mathbf{p}) + \nabla\mathbf{f}\right] - \frac{\beta}{8m}\left[2\boldsymbol{\Sigma}(\boldsymbol{\Phi}\times\mathbf{p}) + \nabla\boldsymbol{\Phi}\right]. \qquad (26)$$

The last summand in (26), instead of direct interaction between spin and gravity $\left(\frac{1}{2}\boldsymbol{\Sigma}\boldsymbol{\Phi}\right)$, describes the spin-orbital and contact interaction of a Dirac particle similarly to the interaction with an electromagnetic field [17].

Note that all the three Hamiltonians (24), (25), (26), are physically equivalent, because they are related to each other by unitary transformations. However, for the quasi-classical interpretation of Hamiltonian terms, one should use the Foldy-Wouthuysen representation [18], [19].

Example 5.

The self-conjugate Hamiltonian in an arbitrary-strength Kerr field derived in [8] differs significantly from Chandrasekhar's Hamiltonian [20], obtained by the Penrose-Newman method [21]. However, after the transformation of Chandrasekhar's Hamiltonian to the $\eta$- representation, we can find that the resultant self-conjugate Hamiltonian is related to the Hamiltonian in [8] by a unitary transformation. Consequently, both Hamiltonians are physically equivalent.

In the general case, the expression for the operator $\eta$ is complex and cumbersome. In the absence of rotation (Schwarzschild field) the operator $\eta$ is diagonal and has the following form:

$$\eta = diag\left[\left(1-\frac{r_0}{r}\right)^{-\frac{1}{2}},1,1,\left(1-\frac{r_0}{r}\right)^{-\frac{1}{2}}\right]. \qquad (27)$$

Now we consider the examples given in the last work of Arminjon [3], in which he demonstrates the non-uniqueness (in his opinion) of the Dirac theory even in the flat Minkowski space.

Example 6.

Arminjon considers a flat Minkowski space, $(t',x',y',z')$, with a free Dirac Hamiltonian

$$H' = \boldsymbol{\alpha}'\mathbf{p}' + \beta'm. \qquad (28)$$

Then he considers a set of other time-dependent Dirac matrices

$$\begin{aligned}\beta &= \beta' \\ \alpha^1 &= \alpha'^1 \cos\omega t - \alpha'^2 \sin\omega t \\ \alpha^1 &= \alpha'^1 \cos\omega t + \alpha'^2 \sin\omega t \\ \alpha^3 &= \alpha'^3.\end{aligned} \qquad (29)$$

As a result, for the new tetrads leading to the set of matrices $\alpha^k$ (29), a new Hamiltonian is obtained:

$$H = \boldsymbol{\alpha}\mathbf{p}' + \beta m - \frac{\omega}{2}\Sigma'^3, \qquad (30)$$

where $\Sigma'^3 = i\alpha'^1\alpha'^2 = i\alpha^1\alpha^2 = \Sigma^3$.

Comparing (28), (30) Arminjon [3] again concludes that the Dirac theory is non-unique and raises the question of physical relevance of direct spin-rotation coupling: $-\frac{\omega}{2}\Sigma'^3$ (this term is present in the Hamiltonian (30), but is absent in the Hamiltonian (28)).

Note that the matrices $\alpha^i$ (29) are related to the initial matrices $\alpha'^i$ by a unitary transformation matrix $R(t)$

$$\alpha^i = R\alpha'^i R^+, \qquad (31)$$

where

$$R(t) = e^{\frac{\omega t}{2}\alpha^{\prime 1}\alpha^{\prime 2}}; \quad R^+(t) = e^{-\frac{\omega t}{2}\alpha^{\prime 1}\alpha^{\prime 2}}. \tag{32}$$

Considering that $R(t)$ is time dependent, we see that the Hamiltonians (30) and (28) are related by the unitary transformation

$$H = RH'R^+ - iR\frac{\partial R^+}{\partial t}. \tag{33}$$

Consequently, the Hamiltonians (28) and (30) are physically equivalent[2]. If the free Hamiltonian (28) transforms to the Foldy-Wouthuysen representation, we obtain the known Hamiltonian [17]

$$H_{FW} = \beta\sqrt{m^2 + \mathbf{p}^2}. \tag{34}$$

Hence, the spin-rotation coupling in (30) is not physically relevant. It can manifest itself with a choice of a specific tetrad field, but it has no effect on the magnitude of final physical characteristics of the system under consideration (absolute analogy with direct spin-gravitation coupling in Example 4).

---

[2] The Dirac equations for Hamiltonians $H'$ and $H$ have the following form:

$$i\frac{\partial \psi'}{\partial t} = H'\psi',$$

$$i\frac{\partial \psi}{\partial t} = H\psi.$$

These equations are equivalent to each other, because the wave functions $\psi'$ and $\psi$ are related by the unitary transformation

$$\psi = R\psi'.$$

For example, the second Dirac equation for the wave function $\psi$ can be written as (considering that $R\frac{\partial R^+}{\partial t} = -\frac{\partial R}{\partial t}R^+$ )

$$i\frac{\partial R}{\partial t}\psi' + iR\frac{\partial \psi'}{\partial t} = RH'\psi' + i\frac{\partial R}{\partial t}\psi',$$

i.e.
$$i\frac{\partial \psi'}{\partial t} = H'\psi'.$$

Example 7.

In his work [3], Arminjon also considers a rotating frame of reference:

$$\begin{aligned}
t &= t' \\
x &= x'\cos \omega t + y'\sin \omega t \\
y &= -x'\sin \omega t + y'\cos \omega t \\
z &= z'.
\end{aligned} \quad (35)$$

The metric corresponding to coordinates (35) is expressed as

$$ds^2 = \left[1 - \omega^2\left(x^2 + y^2\right)\right]dt^2 + 2\omega(ydx - xdy)dt - \left(dx^2 + dy^2 + dz^2\right). \quad (36)$$

In (36), to ensure that $g_{00} > 0$, the condition $\omega\sqrt{x^2 + y^2} < 1$ should be satisfied. $\gamma$-matrices corresponding to the chosen tetrad field have the form

$$\begin{aligned}
\gamma^0 &= \gamma'^0 \\
\gamma^1 &= \gamma'^1 \cos \omega t + \gamma'^2 \sin \omega t + \gamma'^0 \omega y \\
\gamma^2 &= -\gamma'^1 \sin \omega t + \gamma'^2 \cos \omega t - \gamma'^0 \omega x \\
\gamma^3 &= \gamma'^3.
\end{aligned} \quad (37)$$

As a result, we can obtain a self-conjugate Hamiltonian,

$$H_\omega = \boldsymbol{\alpha}'\mathbf{p}' + \beta m - \omega\left(y\frac{\partial}{\partial x} - x\frac{\partial}{\partial y}\right). \quad (38)$$

With another set of tetrads, Arminjon in [3] obtains the following form of $\gamma$-matrices:

$$\begin{aligned}
\gamma^0_{Ar.} &= \gamma'^0 \\
\gamma^1_{Ar.} &= \gamma'^1 + \gamma'^0 \omega y \\
\gamma^2_{Ar.} &= \gamma'^2 - \gamma'^0 \omega x \\
\gamma^3_{Ar.} &= \gamma'^3.
\end{aligned} \quad (39)$$

The self-conjugate Hamiltonian has the following form:

$$H_{Ar.} = \boldsymbol{\alpha}_{Ar.}\mathbf{p}' + \beta m - i\omega\left(y\frac{\partial}{\partial x} - x\frac{\partial}{\partial y}\right) - \frac{\omega}{2}\Sigma^{'3}. \quad (40)$$

Note that the matrices $\gamma^1, \gamma^2$ in (37) can be written as

$$\begin{aligned}
\gamma^1 &= R^+\gamma'^1 R + \gamma'^0 \omega y \\
\gamma^2 &= R^+\gamma'^2 R - \gamma'^0 \omega x.
\end{aligned} \quad (41)$$

One can see from this that the matrices (39) and (37) are related by the unitary transformation

$$\gamma^\mu_{Ar.} = R\gamma^\mu R^+. \quad (42)$$

The Hamiltonians (40) and (38), similarly to the Hamiltonians (30), (28), are physically equivalent, because they are related by the unitary transformation $R(t)$

$$H_{Ar.} = RH_\omega R^+ - iR\frac{\partial R^+}{\partial t}. \quad (43)$$

Thus, as a result of our consideration, we can draw the following conclusions:

1. The problem of non-uniqueness of the Dirac theory in a curved spacetime does not exist. If treated properly, Dirac Hamiltonians will always determine correct physical characteristics of the systems under consideration irrespective of the choice of tetrads.
2. The spin-rotation coupling for Dirac particles in the context of [3] does not represent a physically relevant quantity. It can manifest itself with a certain choice of tetrads, but the spin-rotation coupling has no effect on the final physical characteristics of the quantum mechanical systems under consideration.